\documentclass[12pt]{iopart}

\usepackage{iopams}
\usepackage[cp1250]{inputenc}
\usepackage{amssymb}
\usepackage{amsmath}
\usepackage{multicol}
\eqnobysec
\begin{document}

\title[Relativistic particles in magnetic field and matter]{On the problem of relativistic particles motion in strong magnetic field and dense matter}

\author{I A Balantsev$^1$, Yu V Popov$^2$ and A I Studenikin$^{1,3}$}

\address{$^1$ Department of Theoretical Physics, Moscow State University, Moscow, Russia}
\address{$^2$ Skobeltsyn Institute of Nuclear Physics,
  Moscow State University, Moscow, Russia}
\address{$^3$ Joint Institute for Nuclear Research, Dubna, Russia}
\eads{\mailto{balantsev@physics.msu.ru}, \mailto{popov@srd.sinp.msu.ru}, \mailto{studenik@srd.sinp.msu.ru}}
\begin{abstract}
We consider a problem of electron motion in different media and magnetic field.
It is shown that in case of nonmoving medium and constant homogenious magnetic field the electron energies are
quantized. We also discuss the general problem of eigenvectors and eigenvalues
of a given class of Hamiltonians. We examine obtained exact solutions
for the particular case of the electron motion in a
rotating neutron star with account for matter and magnetic field
effects. We argue that all of these considerations can be usefull for astrophysical applications.
\end{abstract}

\vspace{2pc}

\submitto{\JPA}

\maketitle

\section{Introduction}
Exact solutions of quantum field equations of motion provide an
effective tool in studies of different phenomena of particle
interactions in high energy physics. They supply with
particular applications in solving problems of charged particles
motion in electromagnetic fields of terrestrial experimental
devices, as well as  in astrophysics and cosmology. Exact
solutions were first applied in quantum electrodynamics for
development the quantum theory of the synchrotron radiation, i.e.
for studies of motion and radiation of the electron in a magnetic
field (see, for instance, \cite{SokTerSynRad68}), and also for
studies of the electrodynamics and weak interaction in different
other configurations of external electromagnetic fields
\cite{RitNik1978}. This method is based on the Furry
representation \cite{FurPR51} in quantum electrodynamics (for more
detailed discussion on this item see \cite{StuAFdB06StuJPA08})
widely used for description of particles interactions in the
presence of external electromagnetic fields. Recently it has been
shown, that the method of exact solutions can be also applied for
the problem of neutrinos and electron motion in presence of dense
matter (see \cite{StuAFdB06StuJPA08} for a review on this topic).
Most pronouncedly this possibility was pointed out in
\cite{StuTerPLB05GriStuTerPLB05GriStuTerG&C05GriStuTerPAN06,
GriStuTer05-06} where the exact solution for the modified Dirac
equation for a neutrino moving in matter was derived and discussed
in details. The corresponding exact solution for an electron
moving in matter was obtained in \cite{Stu06GriShiStuTerTro07-08},
the problem of neutrino propagation in transversally moving matter
was first solved in \cite{GriSavStu07}, and in our recent paper
\cite{BalPopStu09} we considered neutrino propagation in a
rotating matter accounting for the effect of nonzero neutrino
mass.

Here we further develop the method of exact solutions for the
problem of charged leptons propagating in matter and strong
magnetic fields. The paper is organized as follows. In Section 2
we discuss the general form of the modified Dirac equation for an
electron moving in matter and magnetic field, and the
corresponding spin operator is derived in Section 3. In Sections
4-6 the exact wave functions and energy spectrum are obtained. The
general problem of eigenvectors and eigenvalues of a given class
of Hamiltonians are discussed. In conclusion (Section 8) we
examine the exact solutions obtained in Sections 2-6 for the
particular case of the charged particle motion in a rotating
neutron star with account for matter and magnetic field effects.

\section{Modified Dirac Equation for electron moving in matter
and magnetic field} We consider an electron propagating in
nonmoving magnetized medium composed of neutrons and suppose, that
magnetic field is homogeneous and constant. This can be regarded
as the first approach to modelling of an electron propagation
inside a neutron star. For distinctness we consider here the case
of an electron, whereas generalization for other charged particles
is just straightforward. We start with the modified Dirac equation
for the electron wave function exactly accounting for the electron
interaction with matter in the absence of magnetic field
\cite{StuTerPLB05GriStuTerPLB05GriStuTerG&C05GriStuTerPAN06} (see
also \cite{GriStuTer05-06}):
\begin{eqnarray}\label{new} \Big\{
\gamma_{\mu}p^{\mu}+\frac{1}{2}
\gamma_{\mu}(1-4\sin^2\theta_W+\gamma^5)f^{\mu}-m \Big\}\Psi(x)=0.
\end{eqnarray}
This is the most general form of the equation for the electron
wave function in which the effective potential
$V_{\mu}=\frac{1}{2}(1-4\sin^2\theta_W+\gamma^5)f_{\mu}$ includes
both the neutral and charged current interactions of the electron
with the background particles, and which can also account for
effects of matter motion and polarization.

In order to include also an effect of an external electromagnetic
field, we replace in eq. (\ref{new}) the electron momentum
$p^{\mu}$ by the "extended" momentum: $p^{\mu}\rightarrow
p^{\mu}+e_0A^{\mu}$,
\begin{eqnarray}\label{new1} \Big\{
\gamma_{\mu}(p^{\mu}+e_0A^{\mu})+\frac{1}{2}
\gamma_{\mu}(1-4\sin^2\theta_W+\gamma^5)f^{\mu}-m \Big\}\Psi(x)=0,
\end{eqnarray}
where $e_0$ is a module of the electron charge.

Note, that in general case it is not a trivial task to find
solutions of this equation. In what follows we consider the
particular case of constant magnetic field and nonmoving uniform
matter, so that for the electromagnetic field and effective matter
potential we obtain
\begin{eqnarray}
A^{\mu}=(0,-\frac{yB}{2},\frac{xB}{2},0), \quad f^{\mu}=-Gn(1,0,0,0),
\end{eqnarray}
where $G=\frac{G_F}{\sqrt{2}}$, $n$ is matter number density. We rewrite equation (\ref{new1}) in the Hamiltonian form and
get
\numparts
\begin{eqnarray}
\rmi\frac{\partial}{\partial t}\Psi(x)=\hat{H}\Psi(x),\label{Dirac equation_1}\\
\hat{H}=\gamma^0\bgamma(\boldsymbol{p}+e_0\boldsymbol{A})+m\gamma^0+\frac{1}{2}(1-4\sin^2\theta_W+\gamma^5)Gn
\end{eqnarray}
\endnumparts
where $\boldsymbol{A}=(-\frac{yB}{2},\frac{xB}{2},0)$. Using the chiral representation of the $\gamma$-matrices we obtain the Hamiltonian in block matrix form:
\begin{eqnarray}\label{bloc-diagonal hamiltonian_1}\fl\hspace{1cm}
\hat{H}=\begin{pmatrix}
&-\bsigma'(\hat{\boldsymbol{p}}+e_0\boldsymbol{A})+Gn(1-2\sin^2\theta_W) \quad& m\\
&m &\bsigma'(\hat{\boldsymbol{p}}+e_0\boldsymbol{A})-2Gn\sin^2\theta_W
\end{pmatrix}
\end{eqnarray}
where $\bsigma'$ are Pauli matrices. This form of the
Hamiltonian makes quite transparent the solution describing spin
properties of the electron.

\section{Spin operator}
It is obvious from (\ref{bloc-diagonal hamiltonian_1}), that the
longitudinal polarization operator
$\hat{T}^0=\frac{1}{m}\bsigma(\hat{\boldsymbol{p}}+e_0\boldsymbol{A})$
\cite{SokTerSynRad68}, where
$\bsigma=\begin{pmatrix}\bsigma'&0\\0&\bsigma'\end{pmatrix}$
are the Dirac matrices, can be written in the form
\begin{eqnarray}
\hat{T}^0=\begin{pmatrix}\label{matrix form of spin operator}
\frac{1}{m}\bsigma'(\hat{\boldsymbol{p}}+e_0\boldsymbol{A})&0\\
0&\frac{1}{m}\bsigma'(\hat{\boldsymbol{p}}+e_0\boldsymbol{A})
\end{pmatrix}
\end{eqnarray}
and commutes with the Hamiltonian, $[\hat{T}^0,\hat{H}]=0$.
Therefore for any of its eigenvectors the Hamiltonian can be
presented as the matrix:
\begin{eqnarray}\label{matrix of hamiltonian}
\hat{H}=\begin{pmatrix}
-mT^0-2Gn\sin^2\theta_W+Gn&m\\
m&mT^0-2Gn\sin^2\theta_W
\end{pmatrix},
\end{eqnarray}
where $T^0$ is one of the eigenvalues of the spin operator
$\hat{T}^0$. Note, that the matrix (\ref{matrix of hamiltonian})
is still $4\otimes 4$ one, and each its element is actually a
product of the number by $2\otimes 2$ unit matrix. In turn, the
transverse polarization operator does not commute with the
Hamiltonian (\ref{bloc-diagonal hamiltonian_1}) because of the
matter term $Gn$ that breaks block symmetry of the Hamiltonian.

\section{Energy spectrum of electron in matter and magnetic field}

In order to find the electron energy spectrum $p_0$ in
the matter and constant magnetic field, $\hat{H}\Psi=p_0\Psi$, we
should solve the equation
\begin{eqnarray}
\begin{vmatrix}
-mT^0+Gn-\tilde{p}_0&m\\
m&mT^0-\tilde{p}_0
\end{vmatrix}=0,
\end{eqnarray}
where $\tilde{p}_0=p_0+2Gn\sin^2\theta_W$. The solutions can be written in the form
\begin{eqnarray}
p_0=\frac{Gn}{2}-2Gn\sin^2\theta_W+\varepsilon\sqrt{(mT^0-\frac{Gn}{2})^2+m^2},\label{spectrum}
\end{eqnarray}
where $\varepsilon=\pm1$ is the "sign" of the energy.

It is significant to note an interesting feature of the electron
energy spectrum in the magnetized matter following from
(\ref{spectrum}). It is well known, that the energy spectrum of
the electron in the magnetic field is degenerated in respect of
spin quantum number (each electron Landau energy level in the
magnetic field corresponds to both spin orientations). The
presence of the matter (of any non-vanishing density $n\neq0$)
removes the degeneracy. This phenomenon can be attributed to the
parity violation in weak interactions.

Let us emphasize one important relation between $p_0$ and $T_0$
that immediately follows from the spectrum (\ref{spectrum}):
\begin{eqnarray}
(p_0-\frac{Gn}{2}+2Gn\sin^2\theta_W)^2=(mT^0-\frac{Gn}{2})^2+m^2,
\end{eqnarray}
where $T^0$ is one of the eigenvalues of the spin operator
$\hat{T}^0$. We use this relation in Section 7.

\section{The electron wave functions}
Note that we can considerably simplify the problem of finding wave
functions if take into account some obvious facts. The solution of
the equation (\ref{Dirac equation_1}) due to symmetries can
be sought in the form
\begin{eqnarray}\label{psi}
\Psi(t,x,y,z) = {\rme}^{-\rmi p_0 t + \rmi p_3z}
\begin{pmatrix}
\psi_1(x,y)\\
\psi_2(x,y)\\
\psi_3(x,y)\\
\psi_4(x,y)\end{pmatrix}.
\end{eqnarray}
Substituting (\ref{psi}) into  (\ref{Dirac equation_1}) we arrive
at a system of linear equations for the electron wave function
components:
\numparts
\begin{eqnarray}\label{system1}
(Gn-p_3)\psi_1+{\rmi}\left\{\left(\frac{\partial}{\partial x}-
{\rmi}\frac{\partial}{\partial y}\right)+\frac{e_0B}{2}(x-{\rmi}y)\right\}\psi_2+m\psi_3=\tilde{p}_0\psi_1,\\
{\rmi}\left\{\left(\frac{\partial}{\partial x}+{\rmi}\frac{\partial}{\partial y}\right)-
\frac{e_0B}{2}(x+{\rmi}y)\right\}\psi_1+(p_3+Gn)\psi_2+m\psi_4=\tilde{p}_0\psi_2,\\
m\psi_1+p_3\psi_3+{\rmi}\left\{\left(\frac{\partial}{\partial x}-
{\rmi}\frac{\partial}{\partial y}\right)+\frac{e_0B}{2}(x-{\rmi}y)\right\}\psi_4=\tilde{p}_0\psi_3,\\
m\psi_2+{\rmi}\left\{\left(\frac{\partial}{\partial x}+{\rmi}\frac{\partial}{\partial y}\right)-
\frac{e_0B}{2}(x+{\rmi}y)\right\}\psi_3-p_3\psi_4=\tilde{p}_0\psi_4,
\label{system4}
\end{eqnarray}
\endnumparts
In the polar coordinates $x+{\rmi} y=r {\rme}^{{\rmi}\phi},
x-{\rmi}y=r{\rme}^{-{\rmi}\phi}$ one obtains
\begin{eqnarray}
\begin{array}{rcl}
\frac{\partial}{\partial x}+{\rmi}\frac{\partial}{\partial
y}={\rme}^{{\rmi}\phi}\left(\frac{\partial}{\partial r}+\frac{{\rmi}}{r}
\frac{\partial}{\partial\phi}\right), \ \ \ \frac{\partial}{\partial
x}-{\rmi}\frac{\partial}{\partial y}=
{\rme}^{-{\rmi}\phi}\left(\frac{\partial}{\partial r}-\frac{{\rmi}}{r}
\frac{\partial}{\partial\phi}\right),
\end{array}
\end{eqnarray}
and the system of equations (\ref{system1}) - (\ref{system4}) reads now
\numparts
\begin{eqnarray}\label{nextsystem1}
(-p_3+Gn)\psi_1+\rmi\rme^{-\rmi\phi}\left\{\frac{\partial}{\partial r}-
\frac{\rmi}{r}\frac{\partial}{\partial \phi}+\frac{e_0B}{2}r\right\}\psi_2+m\psi_3=\tilde{p}_0\psi_1,\\
\rmi{\rme}^{\rmi\phi}\left\{\frac{\partial}{\partial r}+\frac{\rmi}{r}\frac{\partial}{\partial \phi}-
\frac{e_0B}{2}r\right\}\psi_1+(p_3+Gn)\psi_2+m\psi_4=\tilde{p}_0\psi_2,\\
m\psi_1+p_3\psi_3+\rmi\rme^{-\rmi\phi}\left\{\frac{\partial}{\partial r}-
\frac{\rmi}{r}\frac{\partial}{\partial \phi}+\frac{e_0B}{2}r\right\}\psi_4=\tilde{p}_0\psi_3,\\
m\psi_2+\rmi\rme^{\rmi\phi}\left\{\frac{\partial}{\partial r}+\frac{\rmi}{r}\frac{\partial}{\partial \phi}-
\frac{e_0B}{2}r\right\}\psi_3-p_3\psi_4=\tilde{p}_0\psi_4.\label{nextsystem4}
\end{eqnarray}
\endnumparts

It is possible to show that the operator of the total momentum
$J_z=L_z+S_z$, where $L_z=-i\frac{\partial}{\partial\phi}$,
$S_z=\frac{1}{2}\sigma_3$, commutes with the Hamiltonian.
Therefore the solutions can be written in the form
\begin{eqnarray}\left(
\begin{array}{c}
\psi_1\\\psi_2\\\psi_3\\\psi_4\end{array}\right)
=\left(\begin{array}{c}\chi_1(r)\rme^{\rmi(l-1)\phi}\\\rmi\chi_2(r)\rme^{\rmi l\phi}\\
\chi_3(r)\rme^{\rmi(l-1)\phi}\\\rmi\chi_4(r)\rme^{\rmi l\phi}\end{array}\right).\label{form
of solution}
\end{eqnarray}
The solutions are the eigenvectors of the total momentum operator
$J_z$ with the corresponding eigenvalues $l-\frac{1}{2}$. After
substitution of (\ref{form of solution}), the system (\ref{nextsystem1}) - (\ref{nextsystem4}) turns into:
\numparts
\begin{eqnarray}\label{nextnextsystem1}
-(p_3-Gn)\chi_1-\left(\frac{\rmd}{\rmd r}+\frac{l}{r}+\frac{e_0B}{2}r\right)\chi_2+m\chi_3=\tilde{p}_0\chi_1,\\
\left(\frac{\rmd}{\rmd r}-\frac{l-1}{r}-\frac{e_0B}{2}r\right)\chi_1+(p_3+Gn)\chi_2+m\chi_4=\tilde{p}_0\chi_2,\\
m\chi_1+p_3\chi_3+\left(\frac{\rmd}{\rmd r}+\frac{l}{r}+\frac{e_0B}{2}r\right)\chi_4=\tilde{p}_0\chi_3,\\
m\chi_2-\left(\frac{\rmd}{\rmd r}-\frac{l-1}{r}-\frac{e_0B}{2}r\right)\chi_3-p_3\chi_4=\tilde{p}_0\chi_4.\label{nextnextsystem4}
\end{eqnarray}
\endnumparts
Now we define the  "increasing" and "decreasing" operators
\begin{eqnarray}\label{increasing and decreasing operators}
R^+=\frac{\rmd}{\rmd r}-\frac{l-1}{r}-\frac{e_0B}{2}r, \ \ \ \
R^-=\frac{\rmd}{\rmd r}+\frac{l}{r}+\frac{e_0B}{2}r.
\end{eqnarray}
and get the Hamiltonian in the following form,
\begin{eqnarray}\label{matrix form of hamiltonian}
\hat{H}=\begin{pmatrix}
-p_3+Gn&-R^-&m&0\\
R^+&p_3+Gn&0&m\\
m&0&p_3&R^-\\
0&m&-R^+&-p_3
\end{pmatrix}-2Gn\sin^2\theta_W\hat{I},
\end{eqnarray}
where $\hat{I}$ is the unit matrix. Note that the derived form of
the Hamiltonian is spectacular transparent that significantly
simplify the problem of getting the explicit form of eigenvalues
of the Hamiltonian (\ref{Dirac equation_1}). For forthcoming
applications we take into consideration properties of operators
$R^+$ and $R^-$:
\numparts
\begin{eqnarray}
R^+\,\mathcal{L}_s^{l-1}\left(\frac{e_0B}{2} r^2\right)=-\sqrt{2e_0B(s+l)}\,\mathcal{L}_s^l\left(\frac{e_0B}{2}
r^2\right), \\  R^-\,\mathcal{L}_s^{l}\left(\frac{e_0B}{2}
r^2\right)=\sqrt{2e_0B(s+l)}\,\mathcal{L}_s^{l-1}\left(\frac{e_0B}{2} r^2\right),
\end{eqnarray}
\endnumparts
where $\mathcal{L}_s^l$ are the Laguerre functions
\cite{SokTerSynRad68}.

The solution of system (\ref{nextnextsystem1}) - (\ref{nextnextsystem4}) (the eigenvector of the
Hamiltonian (\ref{matrix form of hamiltonian})) can be written in
the form
\begin{eqnarray}\label{solution}
\left(
\begin{array}{c}
\chi_1\\ \chi_2\\ \chi_3\\ \chi_4\end{array}\right)=\sqrt{e_0B}\left(
\begin{array}{c}
C_1\mathcal{L}_s^{l-1}\left(\frac{e_0B}{2} r^2\right)\\
C_2\mathcal{L}_s^l\left(\frac{e_0B}{2} r^2\right)\\
C_3\mathcal{L}_s^{l-1}\left(\frac{e_0B}{2} r^2\right)\\
C_4\mathcal{L}_s^l\left(\frac{e_0B}{2} r^2\right)
\end{array}\right)
\end{eqnarray}

Now we can get the eigenvalues of the Hamiltonian (the energy spectrum) and of the spin operator $T^0$,
\begin{eqnarray}\fl\hspace{0.5cm}
p_0=\frac{Gn}{2}-2Gn\sin^2\theta_W+\varepsilon\sqrt{\left(-\frac{Gn}{2}\pm\sqrt{p_3^2+2e_0B(l+s)}\right)^2+m^2},\quad \varepsilon\pm 1,\label{p_0}\\
T^0=\frac{s'}{m}\sqrt{p_3^2+2e_0B(l+s)},\quad s'=\pm 1.\label{T^0}
\end{eqnarray}
It is easy to see, that the spectrum (\ref{spectrum}) obtained
above is in agreement with expressions (\ref{p_0}) and
(\ref{T^0}). From this energy spectrum, it is straightforward that
the well-known energy spectrum in magnetic field (the Landau
levels) is modified by interaction of the electron with matter.
However the radius of the classical orbits corresponding to a
certain level (\ref{p_0}) doesn't depend on the matter density:
\begin{eqnarray}
\langle R\rangle=\int\limits_0^{\infty}\Psi^{+} r\Psi\,d\boldsymbol{r}=\sqrt{\frac{2N}{e_0B}}.
\end{eqnarray}
To conclude this section we would like to note, that the
effect of electron trapping on circular orbits in magnetized
matter exists, and this can be important for astrophysical
applications.

\section{General problem of eigenvectors and eigenvalues of a given class of Hamiltonians}

We argue in this Section, that the problem considered above is a
particular case of more general problem, which concludes in a
retrieval of eigenvectors and eigenvalues of Hamiltonians with a
given general structure. The examples of these particular cases
can be found in papers
\cite{SokTerSynRad68,GriSavStu07,BalPopStu09}. Taking into account
the increasing interest to such Hamiltonians, we expect that their
profound consideration would be important for applications.

\underline{\bf Theorem}. Let us take into account the following
conditions:

1) $\mathcal{H}$ is a Hilbert space with the basis of eigenvectors
$\psi_{\{n\}}$ of given quantum problem for a Dirac equation ( a
matter and magnetic field are included);

2) in this space, the increasing and decreasing operators are
determined
\numparts
\begin{eqnarray}
\hat{a}\psi_{\{n\}}=f_-(\{n\})\psi_{\{n-1\}},\\
\hat{a}^{+}\psi_{\{n-1\}}=f_+(\{n\})\psi_{\{n\}},
\end{eqnarray}
\endnumparts
where $f_-(\{n\})$ and $f_+(\{n\})$ are known functions that
depend on the set of quantum numbers $\{n\}$;

3) the Hamiltonian has one of the following explicit structures
\begin{eqnarray}\fl\hspace{1.5cm}
\hat{H}=
\begin{pmatrix}
m&0&p&\hat{a}\\
0&m&\hat{a}^+&-p\\
p&\hat{a}&-m&0\\
\hat{a}^+&-p&0&-m
\end{pmatrix}
\hspace{1cm}
\mbox{or}
\hspace{1cm}
\hat{H}=
\begin{pmatrix}
-p&\hat{a}&m&0\\
\hat{a}^+&p&0&m\\
m&0&p&-\hat{a}\\
0&m&-\hat{a}^+&-p
\end{pmatrix},\label{general Hamiltonian structure}
\end{eqnarray}
where $m$ is the mass of a particle, $p$ is (often a third)
component of momentum, although here we could use other symbols
for demonstration the general mathematical structure. Note that
the theorem is also valid in other case when a certain class of
replacements within the matrixes (\ref{general Hamiltonian
structure}) are made, for example,
$\hat{a}\leftrightarrow\hat{a}^+$, $\hat{a}\rightarrow \rmi\hat{a}$,
$\hat{a}^+\rightarrow -\rmi\hat{a}^+$, $p\leftrightarrow-p$,
$m\leftrightarrow-m$, etc. Nevertheless, the disposition of
operators $\hat{a}$, $\hat{a}^+$, and zeros in the matrixes
important here. Note that there should be only one operator
($\hat{a}$ or $\hat{a}^+$) and only one zero in any string and any
column. Remarkably, for any string or column the operators and
zeros occupy only even or only odd positions.

In that case the following statements can be proved:

1) the eigenvector of such a Hamiltonian has the following form
\begin{eqnarray}
\Psi_{\{n\}}=
\begin{pmatrix}
C_1\psi_{\{n-1\}}\\
C_2\psi_{\{n\}}\\
C_3\psi_{\{n-1\}}\\
C_4\psi_{\{n\}}
\end{pmatrix}\label{general solution}
\end{eqnarray}
and the equation for the spectrum is (for the left structure in
(\ref{general Hamiltonian structure}))
\begin{eqnarray}
\begin{vmatrix}
m-E&0&p&f_-(\{n\})\\
0&m-E&f_+(\{n\})&-p\\
p&f_-(\{n\})&-m-E&0\\
f_+(\{n\})&-p&0&-m-E
\end{vmatrix}=0\label{general equation for the spectrum}
\end{eqnarray}
The equation for the right structure in (\ref{general Hamiltonian
structure}) can be written in analogous way. The spectrum can be
obtained explicitly in the form
\begin{eqnarray}
E=\varepsilon\sqrt{m^2+p^2+f_+(\{n\})f_-(\{n\})},\hspace{1cm}\varepsilon=\pm
1;\label{general spectrum}
\end{eqnarray}

2) the spin operator (one of the possible variants) can be
constructed from the Hamiltonian. It is a matrix of operators,
zeros, integrals of motion and other parameters ($m$). A structure
of the spin operator corresponding to the Hamiltonians in
(\ref{general Hamiltonian structure}) can take one of the
following variants:
\begin{eqnarray}
\hat{S}=
\begin{pmatrix}
\ast&0&\ast&\hat{a}\\
0&\ast&\hat{a}^+&\ast\\
\ast&\hat{a}&\ast&0\\
\hat{a}^+&\ast&0&\ast
\end{pmatrix}
\hspace{1cm}
\mbox{or}
\hspace{1cm}
\hat{S}=
\begin{pmatrix}
\ast&\hat{a}&\ast&0\\
\hat{a}^+&\ast&0&\ast\\
\ast&0&\ast&\hat{a}\\
0&\ast&\hat{a}^+&\ast
\end{pmatrix},\label{general structure of spin operator}
\end{eqnarray}
where all free positions $\ast$  should be filled in using the
main condition $[S,H]=0$. Of course, some refinement of structure
of spin operator (or any of its blocks) is possible, for example,
$\hat{a}\rightarrow c\hat{a}$, $\hat{a}^+\rightarrow
c^*\hat{a}^+$, where $c$ is a complex number, $|c|=1$. Note that
the main part of known spin operators (see, for example
\cite{SokTerSynRad68}) has such a structure.

\underline{\bf The proof of the theorem}

1. Let us substitude the solution (\ref{general solution}) into
equation $\hat{H}\psi=E\psi$  where $\hat{H}$ is given by one of
the explicit structures (\ref{general Hamiltonian structure}). The
homogenious system of linear equations is now obtained. The
equation (\ref{general equation for the spectrum}) is a criterion
of  its nonzero solution, and it is the equation for energy
spectrum. To obtain the spectrum (\ref{general spectrum}) we
should solve this equation, what is trivial.

2. Let us demonstrate now that the structure (\ref{general
structure of spin operator}) of spin operator is adequate to the
considered model. Here we study a particular case that similar to
one presented in Section 3, other cases can be solved in an
analogous way. We take into consideration that the Hamiltonian $H$
is given by structure the right-handed part of (\ref{general
Hamiltonian structure}) and the spin operator we can find in the
following form
\begin{eqnarray}
\hat{S}=
\begin{pmatrix}
s_1&\hat{a}&s_5&0\\
\hat{a}^+&s_2&0&s_6\\
s_5&0&s_3&\hat{a}\\
0&s_6&\hat{a}^+&s_4
\end{pmatrix},
\end{eqnarray}
We should check the main condition $[S,H]=0$. After  substitution
the specified expressions for $H$ and $S$ into this equation we
obtain a system of linear equations for finding the coefficients
$s_i$:
\numparts
\begin{eqnarray}
s_2-s_1=2p;\\
s_4-s_3=2p;\\
s_5+s_6=0;\\
m(s_1-s_3)+2ps_5=0;\\
m(s_2-s_4)-2ps_6=0.
\end{eqnarray}
\endnumparts
One of the solutions can be taken in the  form $s_1=s_3=-p$,
$s_2=s_4=p$, $s_5=s_6=0$, and the corresponding spin operator can
be presented as follows
\begin{eqnarray}
\hat{S}=
\begin{pmatrix}
-p&\hat{a}&0&0\\
\hat{a}^+&p&0&0\\
0&0&-p&\hat{a}\\
0&0&\hat{a}^+&p
\end{pmatrix},
\end{eqnarray}
We see that obtained spin operator is similar to one used in
Section 3.

{\it Remark 1.} The symmetry of the Hamiltonian can be violated as
described in Sections 1-4. There we deal with the Hamiltonian that
differs from (\ref{general Hamiltonian structure}, right) by the
term $B\gamma^5$, and $B$ is a constant. In that case, the spin
operator can be constructed in the form
\begin{eqnarray}
\hat{S}=
\begin{pmatrix}
\ast&\hat{a}&0&0\\
\hat{a}^+&\ast&0&0\\
0&0&\ast&\hat{a}\\
0&0&\hat{a}^+&\ast
\end{pmatrix}\label{structure of block spin operator}
\end{eqnarray}
and the spectrum depends on spin quantum number similar to eq.
(\ref{p_0}). Note that the spin operator (\ref{matrix form of spin
operator}) is similar to that in eq. (\ref{structure of block spin
operator}).

{\it Remark 2.} The theorem can be
generalized to the Hamiltonians of more complicated structures
namely
\begin{eqnarray}\fl\hspace{1.5cm}
\hat{H}=
\begin{pmatrix}
h_{11}&0&h_{13}&\hat{a}\\
0&h_{22}&\hat{a}^+&h_{24}\\
h_{13}&\hat{a}&h_{33}&0\\
\hat{a}^+&h_{24}&0&h_{44}
\end{pmatrix}
\hspace{1cm}
\mbox{or}
\hspace{1cm}
\hat{H}=
\begin{pmatrix}
h_{11}&\hat{a}&h_{13}&0\\
\hat{a}^+&h_{22}&0&h_{24}\\
h_{13}&0&h_{33}&\hat{a}\\
0&h_{24}&\hat{a}^+&h_{44}
\end{pmatrix}.\label{general Hamiltonian structure more complicated}
\end{eqnarray}
The equation for the spectrum for this model (e.g., for the
left side of eq. (\ref{general Hamiltonian structure})) is
\begin{eqnarray}
\begin{vmatrix}
h_{11}-E&0&h_{13}&f_-((n))\\
0&h_{22}-E&f_+((n))&h_{24}\\
h_{13}&f_-((n))&h_{33}-E&0\\
f_+((n))&h_{24}&0&h_{44}-E
\end{vmatrix}=0
\end{eqnarray}
However, to solve this equation is not a simple task.

{\it Remark 3.} It is not obvious from the matrix form of the
Hamiltonian (\ref{matrix form of hamiltonian}), that it is a
self-conjugate operator. To prove this, the properties of
"increasing" and "decreasing" operators should be considered more
carefully. A sequenced collection of functions
$F_s^l=\sqrt{|e|B}\mathcal{L}_s^l\left(\frac{|e|B}{2} r^2\right)$
constitute a basis in the Hilbert space with scalar product
defined as
\begin{eqnarray}
\langle F_{s'}^l, F_{s''}^l\rangle=\int_0^\infty F_{s'}^lF_{s''}^lr\,\rmd r.
\end{eqnarray}
So that, we get for each $s$ and $l$
\begin{eqnarray}
\langle F_s^{l-1}, R^- F_s^l\rangle=-\langle F_s^l, R^+ F_s^{l-1}\rangle.
\end{eqnarray}
Hence for operators (\ref{increasing and decreasing operators}) we obtain
$(R^-)^*=-R^+$ and $(R^+)^*=-R^-$, where symbol $*$ implies
Hermitian conjugation of operators. From
the matrix form of the Hamiltonian (\ref{matrix form of
hamiltonian}) we see that $H^*=H$. So that we can use the theorem
for this case.

Note that the increasing and decreasing operators can be used
effectively in this problem because we need in only one
basis $\mathcal{L}_s^l$ for constructing the spinor
(\ref{solution}). In more complicated models, we seem should take
two or even more different basis functions for constructing a
solution.

\section{Spin coefficients}

We go back now to the Hamiltonian (\ref{matrix of
hamiltonian}). In order to find its eigenvectors, we consider the
following chain of transformations
\begin{eqnarray}
(-mT^0+Gn-\tilde{p}_0)C_1+mC_3=0,
\end{eqnarray}
\begin{eqnarray}
(mT^0-\frac{Gn}{2}+\tilde{p}_0-\frac{Gn}{2})C_1=mC_3,\nonumber\\
\left(\tilde{p}_0-\frac{Gn}{2}\right)\left(1+\frac{mT^0-\frac{Gn}{2}}{\tilde{p}_0-\frac{Gn}{2}}\right)C_1=
\left|\tilde{p}_0-\frac{Gn}{2}\right|\sqrt{1-\frac{(mT^0-\frac{Gn}{2})^2}{(\tilde{p}_0-\frac{Gn}{2})^2}}C_3,\nonumber\\
\varepsilon\sqrt{1+\frac{mT^0-\frac{Gn}{2}}{\tilde{p}_0-\frac{Gn}{2}}}C_1=\sqrt{1-\frac{mT^0-\frac{Gn}{2}}{\tilde{p}_0-\frac{Gn}{2}}}C_3,
\end{eqnarray}
Hence, we obtain
\begin{eqnarray}
C_1=\sqrt{1-\frac{mT^0-\frac{Gn}{2}}{\tilde{p}_0-\frac{Gn}{2}}}A,\quad C_3=\varepsilon\sqrt{1+\frac{mT^0-\frac{Gn}{2}}{\tilde{p}_0-\frac{Gn}{2}}}A.
\end{eqnarray}
In the same way we obtain the following expressions for $C_2$ and $C_4$:
\begin{eqnarray}
C_2=\sqrt{1-\frac{mT^0-\frac{Gn}{2}}{\tilde{p}_0-\frac{Gn}{2}}}B,\quad C_4=\varepsilon\sqrt{1+\frac{mT^0-\frac{Gn}{2}}{\tilde{p}_0-\frac{Gn}{2}}}B,
\end{eqnarray}
where $A$ and $B$ are new constants.

Getting of the relation between $C_1$ and $C_2$ (and between $C_3$
and $C_4$) is similar. Namely, we should use the equations
\begin{eqnarray}
(p_3-mT^0)C_{1,3}+\sqrt{2e_0B(l+s)}C_{2,4}=0.
\end{eqnarray}
We finally get
\begin{eqnarray}
A=\sqrt{1+\frac{p_3}{mT^0}}C,\quad B=s\sqrt{1-\frac{p_3}{mT^0}}C,
\end{eqnarray}
with the only coefficient $C$ which has to be defined from the
normalization condition $C_1^2+C_2^2+C_3^2+C_4^2=1$. We obtain
$C=\frac{1}{2}$. Finally, we obtain the wave function:
\begin{eqnarray}\label{exact solution}
\Psi(t,x,y,z) = \rme^{-\rmi p_0 t}\frac{1}{\sqrt{L}}\mathrm{e}^{\rmi p_3z}\sqrt{\frac{e_0B}{2\pi}}\begin{pmatrix}
C_1\mathcal{L}_s^{l-1}\left(\frac{e_0B}{2} r^2\right)\mathrm{e}^{\rmi(l-1)\phi}\\
\rmi C_2\mathcal{L}_s^l\left(\frac{e_0B}{2} r^2\right)\mathrm{e}^{\rmi l\phi}\\
C_3\mathcal{L}_s^{l-1}\left(\frac{e_0B}{2} r^2\right)\mathrm{e}^{\rmi(l-1)\phi}\\
\rmi C_4\mathcal{L}_s^l\left(\frac{e_0B}{2} r^2\right)\mathrm{e}^{\rmi l\phi}
\end{pmatrix},
\end{eqnarray}
where
\begin{eqnarray}\fl
C_1=\frac{1}{2}\sqrt{1-\frac{mT^0-\frac{Gn}{2}}{\tilde{p}_0-\frac{Gn}{2}}}\sqrt{1+\frac{p_3}{mT^0}},\qquad
C_2=\frac{s'}{2}\sqrt{1-\frac{mT^0-\frac{Gn}{2}}{\tilde{p}_0-\frac{Gn}{2}}}\sqrt{1-\frac{p_3}{mT^0}},\label{C_1 and C_2}\\\fl
C_3=\frac{\varepsilon}{2}\sqrt{1+\frac{mT^0-\frac{Gn}{2}}{\tilde{p}_0-\frac{Gn}{2}}}\sqrt{1+\frac{p_3}{mT^0}},\qquad
C_4=\frac{s'\varepsilon}{2}\sqrt{1+\frac{mT^0-\frac{Gn}{2}}{\tilde{p}_0-\frac{Gn}{2}}}\sqrt{1-\frac{p_3}{mT^0}}\label{coefficients 2}
\end{eqnarray}
and $L$ is a normalizing factor.

Equations (\ref{exact solution})-(\ref{coefficients 2}) represent
the exact solution of (\ref{Dirac equation_1}) with the
Hamiltonian (\ref{bloc-diagonal hamiltonian_1}) that describes the
electron moving in matter and magnetic field. Note that in the
case $n=0$, these formulas are reduced to well-known solutions for
the electron wave functions in a constant homogeneous magnetic
field \cite{SokTerSynRad68}.

\section{Application}
It is important to point out, that the obtained exact solution for
the electron motion in matter and magnetic field can be used as
the first approximation of description of particles moving in an
external environment of more complicated configuration. As an
example, we demonstrate how the problem of an electron (or another
charged particle) motion in a rotating matter with magnetic field
can be solved. This problem is of interest in different
astrophysical contexts.

If the angular velocity $\omega$ is small compare to the magnetic
field, we can calculate the spectrum using a standard perturbation
theory with the small parameter $\frac{Gn\omega}{e_0B}\ll1$. If,
for example, we choose for the matter density, angular velocity
and magnetic field the values peculiar for a rotating neutron star
($n=10^{37} sm^{-3}=7.72\cdot 10^{22} (eV)^3,\, \omega=2\pi\cdot
10^3 s^{-1}=2\pi\cdot 0.66\cdot10^{-12} eV,\, B=10^{10}Gs =
7\cdot10^8 (eV)^2$), then the parameter is really small,
\begin{eqnarray}
\frac{Gn\omega}{e_0B}=6.3\cdot10^{-20}\ll1.
\end{eqnarray}
Now we can take the spectrum and wave functions found above (i.e.,
without rotation) as the lowest order of perturbation series and
find the correction term.

We consider the particular case of constant magnetic field and
rotating uniform matter so that the electromagnetic field and effective
matter potential are given by
\begin{eqnarray}
A^{\mu}=(0,-\frac{yB}{2},\frac{xB}{2},0), \quad f^{\mu}=-Gn(1,-\omega y,\omega x,0),
\end{eqnarray}
where $G=\frac{G_F}{\sqrt{2}}$. Let us rewrite eq. (\ref{new1}) in
the Hamiltonian form
\begin{eqnarray}\fl\hspace{1cm}
\rmi\frac{\partial}{\partial t}\Psi(x)=\hat{H}\Psi(x),\label{Dirac equation}\\\fl\hspace{1cm}
\hat{H}=\gamma^0\bgamma(\boldsymbol{p}+e_0\boldsymbol{A})+m\gamma^0+\frac{Gn}{2}(1-4\sin^2\theta_W+\gamma^5)+\\+
\frac{Gn}{2}\gamma^0\gamma^1(1-4\sin^2\theta_W+\gamma^5)\omega y-\frac{Gn}{2}\gamma^0\gamma^2(1-4\sin^2\theta_W+\gamma^5)\omega x,\nonumber
\end{eqnarray}
with $\vec{A}=(-\frac{yB}{2},\frac{xB}{2},0)$. Using again
the chiral representation of the $\gamma$-matrices we obtain the
Hamiltonian in block-matrix form:
\begin{eqnarray}\label{bloc-diagonal hamiltonian}\fl\hspace{2cm}
\hat{H}=\begin{pmatrix}
&-\bsigma(\hat{\boldsymbol{p}}+e_0\boldsymbol{A})+Gn \quad& m\\
&m &\bsigma(\hat{\boldsymbol{p}}+e_0\boldsymbol{A})
\end{pmatrix}-\nonumber\\\hspace{3cm}-2Gn\sin^2\theta_W-Gn\omega
\begin{pmatrix}
&\sigma_1 y-\sigma_2 x& 0\\
&0 &0
\end{pmatrix}.
\end{eqnarray}

It is obvious from (\ref{bloc-diagonal hamiltonian}), that
$H=H_0+H_1$, where $H_1$ is the last term of (\ref{bloc-diagonal
hamiltonian}) and  takes the following form in the polar
coordinates
\begin{eqnarray}
H_1=\begin{pmatrix}
0&-\rmi\rho r\rme^{-\rmi\phi}&0&0\\
\rmi\rho r\rme^{\rmi\phi}&0&0&0\\
0&0&0&0\\
0&0&0&0
\end{pmatrix},\quad \rho=Gn\omega.
\end{eqnarray}
So, we get for the first correction to the energy spectrum
(\ref{spectrum}) of the electron
\begin{eqnarray}\fl
\Delta
p_0^N=\int\Psi_N^*H_1\Psi_N\,dV=\sqrt{2e_0B}C_1C_2\frac{2Gn\omega}{e_0B}\int\limits_0^{\infty}\mathcal{L}_{N-l}^{l-1}(\xi)\sqrt{\xi}
\mathcal{L}_{N-l}^{l}(\xi)\,d\xi,
\end{eqnarray}
where $C_1$ and $C_2$ are the coefficients from eq. (\ref{C_1 and
C_2}). The integral can be calculated explicitly, and finally we
get
\begin{eqnarray}
\Delta p_0^N=\sqrt{2e_0B}C_1C_2\frac{2Gn\omega}{e_0B}\sqrt{N}=\nonumber\\\hspace{3cm}=2Gn\omega C_1C_2\sqrt{\frac{2N}{e_0B}}=2GnC_1C_2\omega\langle R\rangle.
\end{eqnarray}

This shift of levels in the energy spectrum depending on the
energy quantum number $N=0, 1, 2...$ leads to a corresponding
shift in a frequency of synchrotron radiation of electron inside
of dense rotating matter, that can be registered.

\section{Conclusion}
In this paper we found a class of exact solutions of the modified
Dirac equation which describes the charged leptons propagating in
uniform matter and strong constant magnetic field. We also pointed
out, how this approach can be generalized to a given class of
Dirac Hamiltonians. Obtained solution for the particular case of
the electron motion in a rotating neutron star with account for
matter and magnetic field effects can be used as the first
approximation in more complicated models. All of these
considerations can be useful for astrophysical applications.

\end{document}